\documentclass[twocolumn,prl]{revtex4}
\usepackage{epsfig}
\begin{document}
\author{M. W. Mitchell, J. S. Lundeen and A. M. Steinberg}
\affiliation{Department of Physics, University of Toronto, 60 
St.~George St., Toronto, ON M5S 1A7, Canada }
\newcommand{\DateWritten}{\today}
\title{
Super-resolving phase measurements with a multi-photon entangled state
}
\newcommand{\FigWidth}{3.25in}
\begin{abstract} 
Using a linear optical elements and post-selection, we construct an 
entangled polarization state of three photons in the same spatial 
mode.  This state is analogous to a ``photon-number path entangled 
state'' and can be used for super-resolving interferometry.  
Measuring a birefringent phase shift, we demonstrate two- and three-fold 
improvements in phase resolution.  
%
%
\end{abstract}



\maketitle

\newcommand{\be}{\begin{equation}}
\newcommand{\ee}{\end{equation}}
\newcommand{\bea}{\begin{eqnarray}}
\newcommand{\eea}{\end{eqnarray}}
\newcommand{\ket}[1]{\left|#1\right>}
\newcommand{\bra}[1]{\left<#1\right|}
\newcommand{\bk}{{\bf k}}
\newcommand{\etal}{{\em et al.}} 
\newcommand{\degree}{$^{\circ}$~}
\newcommand{\mydegree}{^{\circ}}
\newcommand{\bigepsilon}{{\cal E}}
\newcommand{\polket}{}
\newcommand{\pmff}{\pm 45\mydegree}

{
Interference phenomena are ubiquitous in physics, and often form the
basis for the most demanding measurements.  Obvious examples
include Ramsey interferometry in atomic spectroscopy\cite{Ramsey80},
x-ray diffraction in crystallography \cite{Bragg13} and optical
interferometry in gravitational-wave studies
\cite{Barish99,Caron97}.  
It has been known for some time
that entangled states can be used to perform super-sensitive
measurements, for example in optical interferometry or atomic
spectroscopy \cite{Holland93,Bollinger96,Dowling98}.  The idea has
been demonstrated for an entangled state of two photons
\cite{D'Angelo01}, but for larger number of particles it is difficult
to create the necessary multi-particle entangled states
\cite{Sackett00,Rauschenbeutel00,Zhao03}.  Here we experimentally 
demonstrate a technique for
producing a maximally-entangled state from initially non-entangled
photons.  The method can in principle be applied to generate states of
arbitrary photon number, giving arbitrarily large improvement in
measurement resolution \cite{Kok02,Fiurasek02,Pryde03,Hofmann03}.  The
method of state construction requires non-unitary operations, which we
perform using post-selected linear-optics techniques similar to those
used for linear-optics quantum computing \cite{Knill01,Franson02,
Resch02,Mitchell03,O'Brien03}.
}


Our goal is to create the state
\be
\ket{N::0}_{a,b} \equiv \frac{1}{\sqrt{2}}
\left( \ket{N,0}_{a,b} + \ket{0,N}_{a,b} \right)
\ee
which describes two modes $a,b$ in a superposition of distinct Fock states
$\ket{n_{a}=N,n_{b}=0}$ and $\ket{n_{a}=0,n_{b}=N}$.  This state
figures in several metrology proposals, including atomic frequency
measurements \cite{Bollinger96}, interferometry \cite{Holland93,Campos03},
and matter-wave gyroscopes \cite{Dowling98}.  In these 
proposals the particles occupying the modes are atoms 
or photons.  



The advantage for spectroscopy can be seen in this idealization:
We wish to measure a level splitting $H_{ext} =\varepsilon_{ba} 
b^{\dagger}b$ between modes $b$ and 
$a$ using a fixed number of particles $N$ in a fixed time $T$.  We 
could prepare $N$ copies of the single-particle state 
$(\ket{1,0}_{a,b} + \ket{0,1}_{a,b})/\sqrt{2}$ and allow them to evolve
to the state $\ket{\phi} \equiv (\ket{1,0}_{a,b} + \exp[i \phi]
\ket{0,1}_{a,b})/\sqrt{2}$, where $\phi = \varepsilon_{ba}T/\hbar$.
Measurements of $A_{1}\equiv \ket{0,1}\bra{1,0} + \ket{1,0}\bra{0,1}$ 
on this ensemble give $\left<A_{1}\right> =\cos(\phi)$ with
shot-noise limited phase uncertainty, $\Delta\phi= 1/\sqrt{N}$.  In 
contrast, under the same Hamiltonian $\ket{N::0}$ evolves to
$\left( \ket{N,0} + \exp[i N \phi] \ket{0,N} \right)/\sqrt{2}$.  
If we measure the operator ${A}_{N} \equiv \ket{0,N}\bra{N,0} +  \ket{N,0}\bra{0,N}$, 
we find $\left<A_{N}\right> =\cos(N\phi)$.
The dependence on $N\phi$ rather than $\phi$ is
phase {\em super-resolution}: one cycle 
of $\left<A_{N}\right>$ 
implies a smaller change of $\phi$ (or $\varepsilon_{ba}$) 
than one cycle of $\left<A_{1}\right>$.
Phase {\em super-sensitivity}, a reduction of phase uncertainty, is 
also predicted.
A number of schemes have been proposed
\cite{Holland93,Bollinger96,Sanders97, Huelga97,Dowling98,Kok02,Campos03}
to reach the so-called Heisenberg limit $\Delta\phi = 1/N$.
The simplest proposals would measure the operator ${A}_{N}$.  This can
be implemented with coincidence measurements, as
the probability of detecting all $N$ quanta in a mode  
$(a+b)/\sqrt{2}$ is proportional to  $1+\left<A_{N}\right>$.

A related technique, quantum interferometric optical lithography,
proposes using phase super-resolution to write features smaller than
the single-photon diffraction limit.  There the modes $a,b$ are
spatial, with different propagation directions.  A molecule exposed to
both modes and capable of absorbing $N$ photons would, in effect,
perform the measurement of $A_{N}$ as above, with $N$-fold
super-resolution in position.  Using coincidence detection in place of
two-photon absorbers, this principle has been demonstrated for $N=2$
using down-converted pairs\cite{D'Angelo01}.  In that experiment, two
infrared photons showed the same (angular) resolution as the blue pump
photon which generated them, a factor of two improvement over the
resolution of a single infrared photon.  The question remains as to 
whether resolution can be improved beyond that of the photons used to 
generate the entangled state.  Here
we answer that question in the affirmative by constructing a
multi-particle state with greater phase resolution than any of its
single-particle precursors.  The technique could in principle be used
to generate entangled states of arbitrarily large $N$ with 
arbitrarily good resolution.



We prepare the state $\ket{3::0}_{a,b}$  where the modes $a$
and $b$ are the horizontal (H) and vertical (V) polarizations of a single
spatial mode.  The construction of the polarization state is based on
earlier proposals to construct photon-number path entangled states
\cite{Kok02,Fiurasek02,Pryde03}.  A similar technique for polarization 
has recently 
been independently proposed \cite{Hofmann03}.  The key to the 
construction is the fact that $\ket{3::0}_{a,b}$, when 
written in terms of creation operators 
$a^{\dagger}_{a}$ and $a^{\dagger}_{b}$ acting on the vacuum $\ket{0}$, is
\be
\ket{3::0}_{a,b} =
(a^{\dagger}_{a} + a^{\dagger}_{b})
(a^{\dagger}_{a} + e^{i\chi} a^{\dagger}_{b})
(a^{\dagger}_{a} + e^{i2\chi}a^{\dagger}_{b})\ket{0}
\ee
where $\chi = 2 \pi/3$ and normalization has been omitted.  
The terms in parentheses each create a 
particle, but in non-orthogonal states.  If $a$ and $b$ are left 
and right circular polarization, these states describe linear 
polarizations rotated by $60\mydegree$ from each other.  Using 
post-selection, we can put one photon of each polarization into a 
single spatial mode and create $\ket{3::0}_{a,b}$.

%
%
\begin{figure}[h]
\centerline{\epsfig{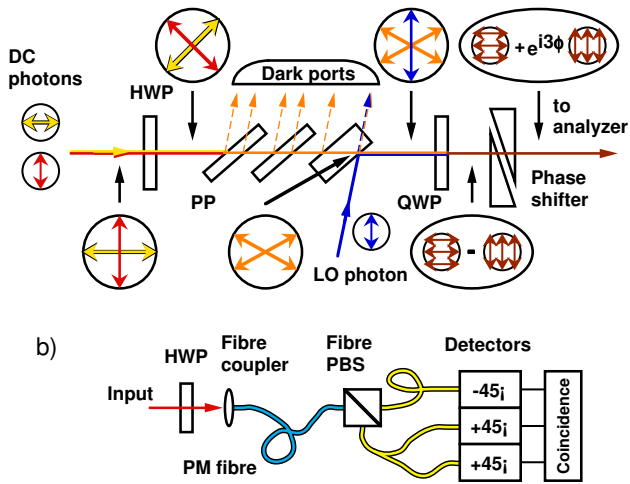}}
\caption{Schematic of production and detection of the state
$\ket{3::0}_{H,V}^{3\phi}$.  Part a) shows the chain of optical 
components and their effect on polarization state, represented in 
encircled figures.  All photons have wavelength
810 nm; colours indicate photon origins and are blended after
operations requiring indistinguishability.  A polarizing
beamsplitter (PBS) puts the DC photons in a single spatial mode
and a half wave-plate (HWP) rotates their polarizations to 
$\pm 45$\degree.  A partial polarizer (PP) transforms the 
polarizations to $\pm 60$\degree if no photons are reflected
into the dark ports.  The LO photon is injected at the
final interface of the partial polarizer.
A quarter wave-plate (QWP) rotates to
$(\ket{3,0}_{H,V}-\ket{0,3}_{H,V})/\sqrt{2}$ and quartz wedges produce
an adjustable phase $3\phi$ between the two components of the state. 
%
b)
analysis in the $\pm 45\mydegree$ polarization basis is
performed with a HWP before a
polarization-maintaining (PM) fibre and a fibre-coupled PBS.  The 
outputs of the fibre PBS are 
channeled to one, two, or three detectors as needed.  
The configuration to detect $\ket{2,1}_{\pmff}$ is shown.
Digital electronics record single detections as well as two- and
three-fold coincidences.  }
\end{figure}

We use two photons from pulsed parametric down-conversion 
plus one laser, or ``local oscillator'' (LO) photon.  Pulses
of $100$ fs duration and 810 nm center wavelength are produced by 
a mode-locked
Ti:sapphire laser and frequency-doubled giving 405 nm pulses with
an average power of $50$ mW. These are gently focused into a 0.5 mm
thick beta-barium borate crystal aligned for Type-II parametric
down-conversion in the ``collapsed cone'' geometry \cite{Takeuchi01}.  The 
down-converted (DC)
photons thus produced are orthogonally polarized.  
A small part of the Ti:sapphire beam is split
off and attenuated to contribute the LO photon.  
These three photons
are
transformed into the state $\ket{3::0}$ and detected by 
polarization-resolved three-fold coincidence 
detection.  The transformation, shown in Figure 1 a), can be understood as a 
sequence of mode combinations.  After the PBS, the DC photons are in the state
$a^{\dagger}_{H}a^{\dagger}_{V}\ket{0}$, where subscripts indicate
polarization.  
A half wave-plate 
rotates this to 
$a^{\dagger}_{+45\mydegree}a^{\dagger}_{-45\mydegree}\ket{0}$ where 
subscripts show linear polarizations measured from vertical.
We perform the first post-selected non-unitary operation by
passing the pair through three plates of BK7 glass near 
the Brewster angle.  The six Brewster-angle interfaces act as a partial
polarizer (PP) with transmission efficiencies of 
$T_{H}\approx 1$, $T_{V} = 1/3$.  By post-selecting cases where no 
photons are reflected, we transform the state as (again without 
normalization)
\bea
a^{\dagger}_{+45\mydegree}a^{\dagger}_{-45\mydegree}\ket{0} &= &
(a^{\dagger}_{H}+a^{\dagger}_{V})(a^{\dagger}_{H}-a^{\dagger}_{V})\ket{0}
\nonumber \\
&\rightarrow&(a^{\dagger}_{H}+\frac{1}{\sqrt{3}}a^{\dagger}_{V})
(a^{\dagger}_{H}-\frac{1}{\sqrt{3}}a^{\dagger}_{V})\ket{0}
\nonumber \\
&=& a^{\dagger}_{+60\mydegree}a^{\dagger}_{-60\mydegree}\ket{0}.
\eea
This operation, 
putting orthogonally polarized photons into non-orthogonal modes,
is non-unitary and requires post-selection.

The DC photons meet the LO photon at the last interface of the PP.  
This interface acts as a beamsplitter putting all three
into the same spatial mode, conditioned on zero photons exiting by the ``dark''
port.   Again, the operation is non-unitary and requires
post-selection.  The LO photon is vertically polarized and the state
thus constructed is $a^{\dagger}_{0\mydegree}a^{\dagger}_{+60\mydegree}
a^{\dagger}_{-60\mydegree}\ket{0}$.  This state
has
six-fold rotational symmetry about the beam propagation direction, 
and thus 
can only contain the states $\ket{3,0}_{L,R}$ and $\ket{0,3}_{L,R}$
where subscripts indicate the circular polarization basis 
\cite{Hofmann03}.  In fact, it is easily
verified that the state is $(\ket{3,0}_{L,R} -
\ket{0,3}_{L,R})/\sqrt{2}$ up to an overall phase.  Finally, we 
convert circular to linear polarizations with a quarter wave-plate, 
giving $(\ket{3,0}_{H,V} - \ket{0,3}_{H,V})/\sqrt{2}$.  Ideally, 
the probability of both post-selections succeeding is 
$ \cos^{4}(\pi/12)/3^{1/6} \approx 72\%$. 

Response to the phase shifter demonstrates phase super-resolution.
Acting on a single photon, the quartz wedges shift by $\phi$ the phase
of the $V$-polarization.  Acting on three photons the phase shift is
tripled: $\ket{3::0}_{H,V}$ becomes $\ket{3::0}^{3\phi}_{H,V}\equiv
(\ket{3,0}_{H,V} + \exp[i3\phi] \ket{0,3}_{H,V})/\sqrt{2}$, where we
have absorbed the negative sign above into the phase factor.  The
$3\phi$ behaviour can be seen in triples detection in the $\pm
45\mydegree$ linear polarization basis.  The rates for detection of
$\ket{3,0}_{\pmff}$ and $\ket{2,1}_{\pmff}$ vary as
$1\pm\cos{(3\phi)}$, respectively.  After passing through an 810 nm wavelength 
filter with a 10 nm passband, the photons enter the polarization analyzer, 
set to detect $\ket{3,0}_{\pmff}$ or $\ket{2,1}_{\pmff}$, as shown in Figure
1 b).


The use of down-converted pairs removes the need for detectors at the
``dark'' ports.  Down-conversion very infrequently produces more than
two photons in a single pulse so we can infer with near-certainty the
absence of photons in the ``dark'' port from the presence of both
photons in the ``bright'' port.  Using a weak coherent state to supply
the third photon, we can make small the probability that more than one
LO photon was present in a triple detection event \cite{Pryde03}.
Thus a single post-selection for three-fold coincidence at the
detectors performs at once the post-selections for both non-unitary
operations.

%

\begin{figure}[h]
\centerline{\epsfig{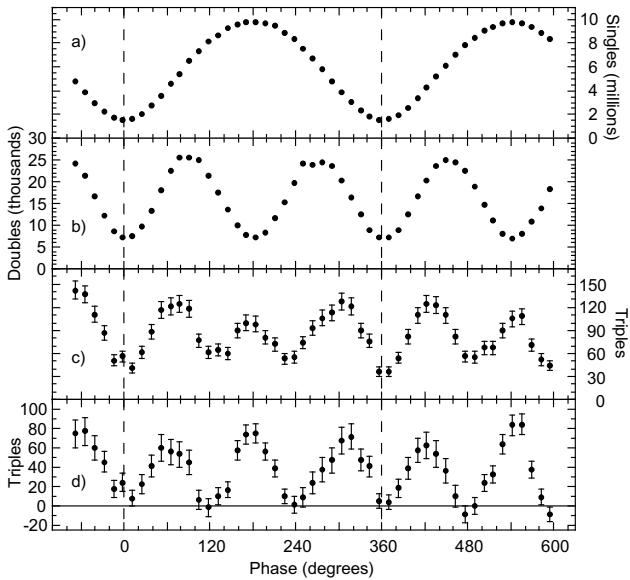}}
\caption{Super-resolving phase measurement with two and three 
photons.  The polarization analyzer is 
configured with one
detector on the -45\degree output and two on +45\degree.  
The input, a combination of down-converted 
and laser photons, is constructed to give the state 
$\ket{3::0}_{H,V}^{3\phi}$ upon post-selection (see text).
All graphs show detections per 30 second counting 
interval as the phase $\phi$ is changed by translating a phase-shifter 
prism.  a) singles detection at -45\degree, i.e., detection of
$\ket{0,1}_{\pmff}$,
shows oscillation with $\phi$.  b) two-fold coincidence detection
of $\ket{1,1}_{\pmff}$ shows 
oscillation with $2\phi$.  c) three-fold coincidence detection of
$\ket{2,1}_{\pmff}$ shows oscillation with $3\phi$.  Error bars 
indicate $\pm \sigma$ statistical uncertainty. d) three-fold 
coincidence after background subtraction.  Error bars show  $\pm \sigma$ 
statistical uncertainty plus a systematic uncertainty in the 
background.  Dashed vertical bars indicate one full cycle.
}
\end{figure}

Figure  2 shows results for detection of multiple polarizations at the 
analyzer.
Intensities of the DC and LO 
sources were adjusted such that 
singles detections are mostly
produced 
by LO photons (about 10:1 ratio vs. DC singles), two-fold coincidences mostly  
by DC pairs (about 5:1 vs. LO accidentals), and three-fold coincidences 
principally by
one LO photon plus one DC pair (about 2:1 vs. accidental triples 
contributions, below).
Thus with a single scan of the phase shifter we can see qualitatively
different behaviours for states of one, two, and three photons.  Figure
2 c) clearly shows oscillation with $3\phi$, as predicted by theory.  
The resolution exceeds that achievable with any single photon in the
experiment.  The 405 nm photons would show oscillation with $2.1 \phi$ 
due to their shorter wavelength and the somewhat larger birefringence 
of quartz at that wavelength.  A
cosine curve fitted to these data shows a visibility of $42 \pm 3 \%$.
This visibility is unambiguous evidence of indistinguishability and
entanglement among the three photons.  A non-entangled state of three
{\em distinguishable} photons could also show three-fold coincidence
oscillation at $3\phi$, but with a maximal visibility of $20\%$.  

Figure
2 d) shows the same triples data after subtraction of background from
accidental triples.  In addition to the signal of interest
from 1 DC pair + 1 LO photon, we also see events from 2 DC pairs,
from 3 LO photons, and from 2 LO photons + 1 DC pair.  We calculate
these backgrounds from independent measurements of single and double
detection rates for the DC and LO sources alone.  Coincidence
background is calculated by the statistics of uncorrelated sources
using a time-window of 12.5 ns, the laser pulse period.
Incoherence of the various contributions is ensured by sweeping the
path length of the LO photon over $\pm 2 \mu m$ during acquisition.
The calculated background has some variation with $\phi$, so it is
important to note that it is qualitatively different than the observed
$3\phi$ signal.  Per 30 second interval, the accidental
background contributes $22\pm 1$ as a constant
component, an average of $23 \pm 1$ oscillating with $2\phi$, 
$4 \pm 1$ with $1\phi$ and $<1$ with $3\phi$.  Here and elsewhere, 
uncertainty in the counting circuitry's dead-time introduces a 
systematic error.



\begin{figure}[h]
\centerline{\epsfig{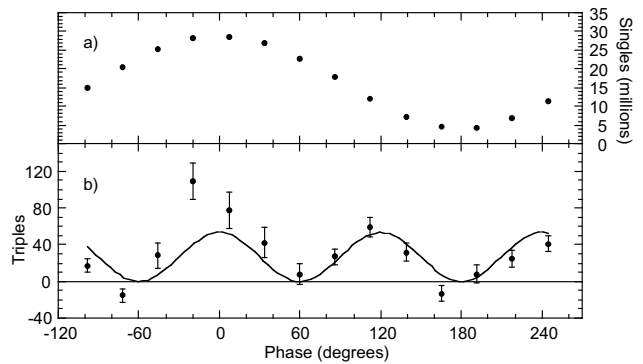}}
\caption{Super-resolving phase measurement with a single detected
polarization.  The polarization analyzer is 
configured to detect $\ket{3,0}_{\pmff}$, i.e., with
three detectors on the +45\degree 
channel.  The input state is the same as in figure 2.
Graphs show detections per 300 second counting 
interval as the phase $\phi$ is changed.
a) singles detection shows oscillation with $\phi$.  b) three-fold 
coincidence detection (after background subtraction) shows 
oscillation with $3\phi$.  
Error bars show $\pm \sigma$ statistical uncertainty plus a 
systematic uncertainty in the  background.  Curve is the expected 
signal $A [1+\cos(3\phi)]$ with $A$ chosen for best fit.
}
\end{figure}

It is also possible to see $3\phi$ behaviour detecting a single 
polarization, as shown in Figure 3.
This measurement corresponds to
the original proposals for atomic spectroscopy and lithography
\cite{Bollinger96,Kok02}.  It gives a far weaker signal, in part
because the maximum overlap of the state $\ket{3::0}^{3\phi}_{H,V}$
with $\ket{3,0}_{\pmff}$ is smaller than with $\ket{2,1}_{\pmff}$.
Also, the chance that all three photons go to distinct detectors (as
needed for coincidence detection) is smaller for $\ket{3,0}_{\pmff}$.   
With these limitations, we are able to see the
$3\phi$ behaviour, but only by subtracting a considerable coincidence
background.

Using linear optical elements and post-selection, we have constructed a
multi-particle entangled state useful for super-resolving phase
measurements.  The demonstrated resolution is not only better than for
a single infrared photon, it is better than could be achieved with any
single photon in the experiment, including the down-conversion pump
photons.  Given the difficulty of generating coherent short-wavelength
sources, this is encouraging for the prospects of proposals such as
quantum-interferometric optical lithography.  Finally, we note that the
construction, which procedes from unentangled photons in distinct
spatial modes to a maximally-entangled state in a single spatial mode,
does not require prior entanglement of the component photons.  As such,
the method is well adapted for use with single-photon-on-demand sources
\cite{Michler00,Solomon01} which promise to be more efficient and scalable
than down-conversion sources.

\section{Acknowledgements}
We thank K. Resch and J. O'Brien for helpful and stimulating
discussions and J. Dowling 
for inspiration.  This
work was supported by the National Science and Engineering Research
Council of Canada, Photonics Research Ontario, the Canadian Institute
for Photonic Innovations and the DARPA-QuIST program (managed by AFOSR
under agreement No.  F49620-01-1- 0468).
%
%

\bibliography{NooNBib}
\bibliographystyle{apsrev}

\end{document}